# Hazardous Asteroids Classification

1st Alvin Buana
*Computer Science and Information Engineering, National Central University*
Zhongli, Taiwan
Alvinbuanaa@gmail.com

2nd Josh Lee
*Computer Science and Information Engineering, National Central University*
Zhongli, Taiwan
joshlee.sesc02@gmail.com

3th Thai Duy Quy
*Computer Science and Information Engineering, National Central University*
Zhongli, Taiwan
quytd@dlu.edu.vn

4st Rakha Asyrofi
*Computer Science and Information Engineering, National Central University*
Zhongli, Taiwan
asyrofi@hangtuah.ac.id

*Abstract*—Hazardous asteroid has been one of the concerns for humankind as fallen asteroid on earth could cost a huge impact on the society. Monitoring these objects could help predict future impact events, but such efforts are hindered by the large numbers of objects that pass in the Earth's vicinity. The aim of this project is to use machine learning and deep learning to accurately classify hazardous asteroids. A total of ten methods which consist of five machine learning algorithms and five deep learning models are trained and evaluated to find the suitable model that solves the issue. We experiment on two datasets, one from Kaggle and one we extracted from a web service called NeoWS which is a RESTful web service from NASA that provides information about near earth asteroids, it updates every day. In overall, the model is tested on two datasets with different features to find the most accurate model to perform the classification

## I. Introduction

The use of artificial intelligence has been significantly used in solving real world cases. It is now essential to implement AI algorithms to make tools in space science. Asteroids, objects that orbit within the Solar system, have been seemingly scattered infinitely across the entire space. It is common to notice them on the asteroid belt between Jupiter and Mars. However, the movement of these rocks could be altered due to the collision of other objects. Some of the results of the collision have a possibility of leading a path to Earth. Furthermore, these asteroids could be harmful to humankind if these asteroids come to the Earth's atmosphere. An asteroid struck Earth's atmosphere on June 30, 1908, and burst in the skies over Siberia. Witnesses in the remote area claimed to have heard a loud explosion and seen a fireball. They also reported trees that were blown over for miles and large forest fires [1]. On 15 February 2013, an asteroid twenty meters wide hit Chelyabinsk, Russia. The shockwave from the explosion was massive which injured more than a thousand people and destroyed buildings across six cities [2]. To prevent the hazardous asteroid from becoming a potential threat, data from previous events are needed to be studied more. However, is it beyond human capability to study thousands of data to predict the future outcome of potential hazardous asteroids. Thus, the existence of artificial intelligence would help us to determine whether an asteroid is potentially dangerous or the opposite. In the field of asteroid classification, other research studies have proposed other methods that would helped classifying hazardous asteroids. In [3], the authors used a data from an official website of Jet Propulsion Laboratory of California Institute of Technology which is an Organization under NASA. They use 16 features from 40 features and also created a benchmark of models. The result is random forest having the best model to perform. In another paper [4], they use PHAC hybrid classifier model that included a deep autoencoder trained upon the vast collection of non-hazardous asteroid data. Their approach successfully achieves 90% accuracy 99% precision and an 80% recall score. In [5], the authors used the most well-known models in Deep Learning such as Inception, Resnet, Xception… to changes to CNN and training it. The accuracy they can reach over 99%.

The main goal of the present project is to review the effectiveness of Machine Learning and Deep Learning methods at the task of classifying astronomical objects, specifically asteroids. We implement five different algorithms on machine learning and five models on deep learning on two different datasets to determine which model that is suitable for classifying hazardous asteroids. The reason to use deep learning in this project is because deep learning has helped solving thousands of problems including classification. Also, the use of deep learning in this project will provide us some results to know if the performance of deep learning models could outperform the machine learning models in asteroid classification. We use two different datasets that has different features and it will be used to test the model in different dataset.

This report is structured as follows: Section 2 presents some related knowledge on the presented topic, while Section 3 presents our methodology to preprocessing data applying to our proposal. The next sections describe the result and some discussion. The last section concludes the ideas presented in this paper.

## II. Preliminary

To evaluation the efficient of the proposal, we use 5 machine learning algorithms and 5 deep learning models.

### A. Machine Leaning algorithms

- **Logistic Regression**: Logistic Regression is statistical model is often used for classification and predictive analytics [6]. In most cases, Logistic Regression is implemented for binary classification due to the fact that the model uses a sigmoid function where the output is ranged between 0 and 1. This logistic function is represented by the following formulas 1.

$$f(x) = \frac{L}{1+e^{-k(x-x_0)}} \quad (1)$$

- **K-Nearest Neighbor**: The k-nearest neighbors algorithm, also known as KNN or K-NN, is a non-parametric, supervised learning classifier, which uses proximity to make classifications or predictions about the grouping of an individual data point [7]. While it can be used for either regression or classification problems, it is typically used as a classification algorithm, working off the assumption that similar points can be found near one another.

- **Support Vector Machine**: Support vector machines (SVM) was firstly introduced for classification and nonlinear function estimation. It is a method that was created in 1964 and the method proposes a kernel trick to maximum-margin hyperplanes. The Algorithm focuses on generating the best suited hyperplane from the given train dataset so that the new data can be shown and classified directly [8]. The SVM's accuracy is proportional to the size of the dataset. Overfitting occurs as the size and number of attributes increase, reducing accuracy.

- **Random Forest**: Random forest is similar to a decision tree algorithm. However, Random Forest uses a bagging method that is consist of several decision tree that will be trained together. The random forest will then use the average decision of the trees that will be put into a consideration to determine the final decision.

- **Catboost**: CatBoost is a machine learning model that uses decision tree on both classification and regression. CatBoost uses a combination of ordered boosting, prediction shift, and gradient-based optimization that can handle categorical data effectively [10]. This model can save a lot of time in preprocessing since it can handle numerical, categorical, as well as text data that should have been handled in the preprocessing data. The model is also considered as a gradient boosting model or ensemble model which means that the model will be able to perform a series of decision trees which will increase the performance of the model. The main difference between this model and random forest tree is the flexibility of CatBoost model compare to the random forest tree.

B. *Deep Learning model*

- **Multilayer Perceptron (MLP)**: MLP is a type of artificial neural network and a fundamental form of deep learning model. It consists of multiple layers of nodes (or neurons) arranged in a feedforward fashion, meaning that information flows from the input layer through one or more hidden layers to the output layer without any cycles or loops. Common activation functions include sigmoid, hyperbolic tangent (tanh), and rectified linear unit (ReLU).

- **Deep Neural Network (DNN)**: A deep neural network (DNN) is a type of artificial neural network with multiple layers between the input and output layers. The term "deep" refers to the depth of the network, which is determined by the number of hidden layers it contains (Fig. 1) These networks are designed to automatically learn hierarchical representations of data, enabling them to capture complex patterns and features. Common activation functions include sigmoid, hyperbolic tangent (tanh), and rectified linear unit (ReLU)

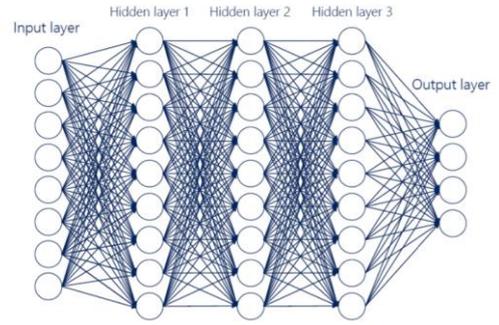

Fig. 1. An example of the Deep Neural Network

- **Convolutional Neural Network (CNN)**: CNN designed for tasks involving visual data, such as image recognition, object detection, and image classification. CNNs are particularly effective in capturing spatial hierarchies of features within images. CNNs use activation functions like ReLU (Rectified Linear Unit) to introduce non-linearity, enabling the network to learn complex relationships in the data.

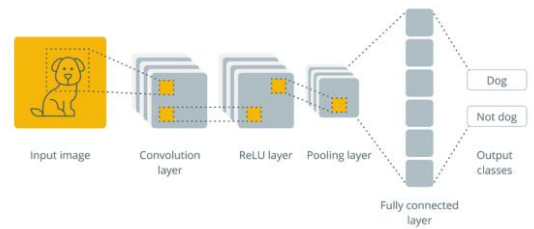

Fig. 2. An example of the Convolutional Neural Network

- **Simple RNN model:** A recurrent neural network (RNN) is a class of artificial neural networks where connections between nodes can create a cycle, allowing output from some nodes to affect subsequent input to the same nodes. This allows it to exhibit temporal dynamic behavior (Fig. 3).

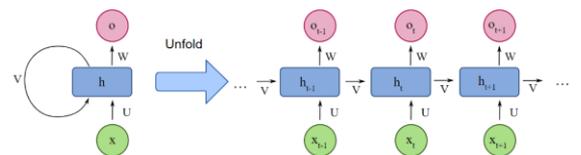

Fig. 3. An example of Simple Recurrent Neural Network (RNN)

- **Long short-term memory (LSTM):** LSTM is the advanced of the RNN model, it can get feedback connections. Such a recurrent neural network (RNN) can process not only single data points (such as images), but also entire sequences of data (such as speech or video) (Fig. 4).

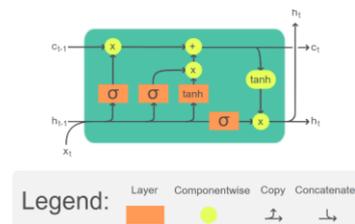

Fig. 4. An example of Long short-term memory (LSTM)

## III. RESEARCH METHODOLOGY

### A. Proposed Methods

In order to build a robust model that can classify the type of asteroids. We have made several methods that could help determine the hazardous asteroids. The proposed methods that we are currently proposing are traditional algorithms which are logistic regression, Support Vector Classifier, KNN, Random Forest, and CatBoost classifier. Meanwhile we are also going to include deep learning algorithms for our comparison which are deep neural network, convolutional 2 Dimensional neural network, multi-layer perceptron classifier, recurrent neural network, and LSTM. Since both data have imbalance labels, we decided to implement bootstrapping and SMOTE to handle the imbalance data. Fig. 5 illustrates the model architecture of our approach.

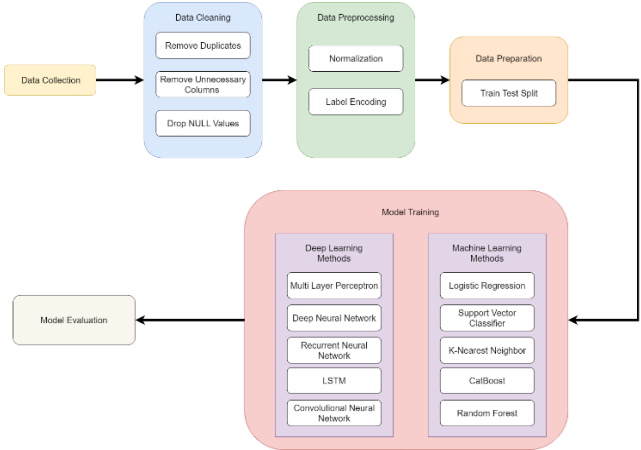

Fig. 5. Proposed Method

### B. Datasets

Our dataset is consisting of two datasets. The first dataset is a public Kaggle dataset called "Asteroid Data" by Mir Sakhawat Hossain. This dataset includes 41 features and 362717 samples. The second dataset is created by our project, we crawl from a web service called Near Object Earth Web Service (NeoWs), name NeoWs [8]. The dataset describes the features of an asteroids such as the physical measurements, the date of the asteroids and the labels to determine whether the asteroid is hazardous or normal to earth. Table I describes the two datasets with the number of features and samples.

TABLE I. DATASETS USING ON PROPOSAL METHOD

| Dataset | File Type | Features | Samples | Link download |
|---------|-----------|----------|---------|---------------|
| Asteroid | csv | 45 | 958524 | https://www.kaggle.com/datasets/sakhawat18/asteroid-dataset |
| NeoWS | csv | 41 | 362717 | https://www.kaggle.com/datasets/alvinb/neows-hazardous-asteroid-dataset |

### C. Preprocessing

Because the datasets include various redundant features, it will be making the algorithm with low accuracy. Thus, we need apply the preprocessing on the data to get more accuracy. The dataset is imported and transformed into a data frame where the data frame will then be cleaned by removing the unnecessary columns and the null values since the dataset have a plethora of null values. Actually, the asteroids on the Solar system is located far from the Earth, so almost the observe asteroid are unbalance. Hence, the dataset will then be put into a balancer either using bootstrapping or SMOTE because the dataset is suffering from imbalance labels with only 13% labels are considered true and 87% labels are considered false on NeoWs dataset. Meanwhile the asteroid dataset on Kaggle have 0.002% labels that are considered true and 99.998% labels that are considered as false. Fig. 6 and Fig. 7 present the both dataset before and after apply an imbalance method.

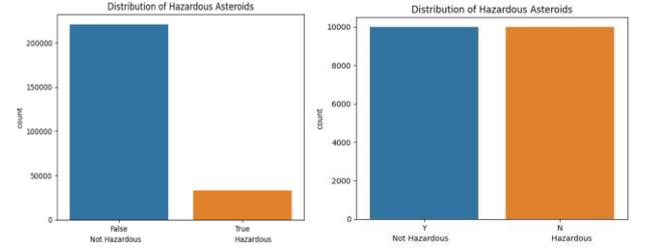

Fig. 6. Distribution of hazardous asteroid on NeoWs dataset before and after apply imbalance algorithm

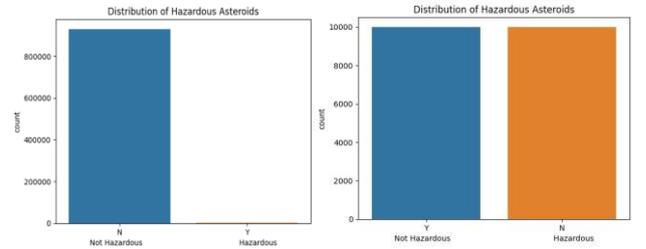

Fig. 6. Distribution of hazardous asteroid on Asteroid dataset before and after apply imbalance algorithm

After setting the balancer on both datasets, the datasets are then normalized with minmax scaler which will scale all of the numerical data into 0 to 1. For categorical data, the label encoder is used to encode all of the categorical labels. The reason to use both of these technique before the modelling benchmark is to simplify the process of the model training and minimize the redundancy.

### D. Data Preparation

In the data preparation, each dataset is divided and shuffled into train and test data where the train data will be used for the model training and the test data will be used for testing the performance of the model. For deep learning models, after the data are split into train and test data, the data are reshaped into the needed input shape for each deep learning models.

### E. Training models

After preprocessing data, in the machine learning models, the data that has been prepared can be put directly into training and testing. For deep learning models, such as CNN or RNN we need transform data to fix with the input model. To get high accuracy, we make one hot encoder for the output and use softmax function as activation output.

In the DNN, we design 4 layers with the number of each layer are 128, 64, 32 and 8 nodes. The output of model only one because we need classification the asteroid is dangerous or not (Fig. 7).

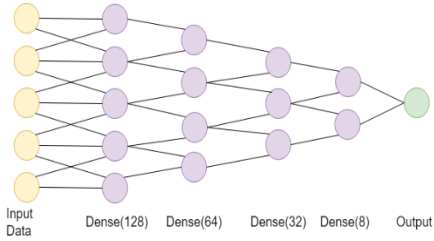

Fig. 7. Deep Neural Network

In the CNN model, we built three Conv2D with the number of each layer are 64, 32, 16. Then we put the output of the convolutional layers to max pooling and apply 32 dense layers before output layer (Fig. 8).

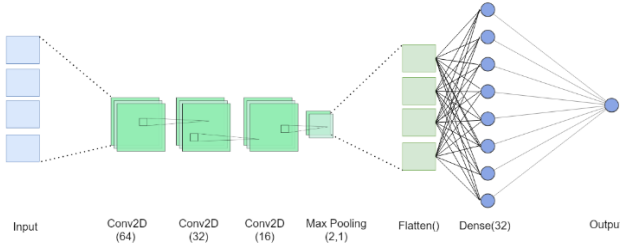

Fig. 8. Convolutional Neural Network

In the RNN and LSTM models, we built the model with 64 nodes for input layers. The time step we use in the both model is 1, the output we make flatten and put to the output (Fig. 9, Fig. 10).

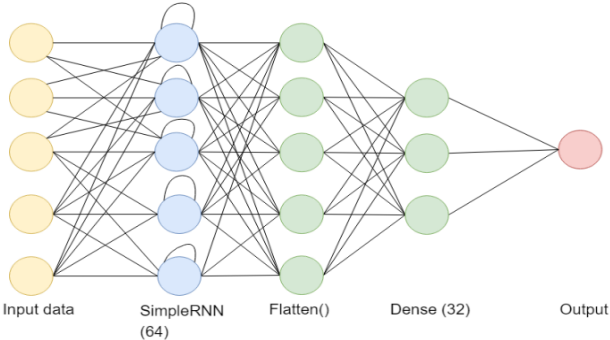

Fig. 8. Recurrent Neural Network

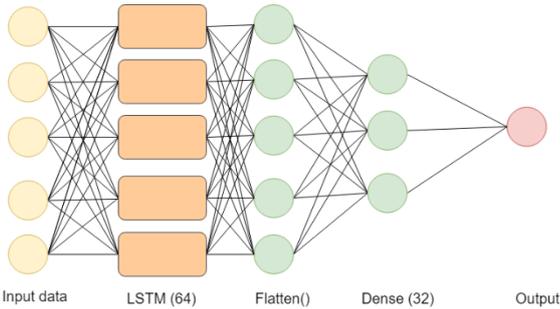

Fig. 8. LSTM model

*F. Evaluation metrics*

One drawback when using accuracy to compare results from different models is that it is heavily affected by class imbalance that refers to an unequal distribution of different classes or categories within a dataset. For our particular dataset, this could be an issue based on the fact that the dataset is imbalanced. For this reason, we do not rely only on accuracy as our main metric.

$$Accuracy = \frac{TP+TN}{TP+FP+TN+FN} \quad (2)$$

where TP is true positives, TN is true negatives, FP is false positives, and FN is false negatives.

Precision is therefore computed as the number of true positives (asteroid objects correctly identified) divided by the sum of true positives and false positives (non-asteroid objects marked incorrectly as asteroid objects)

$$Precision = \frac{TP}{TP+FP} \quad (3)$$

Recall takes into consideration the false negatives and is computed as the number of true positives divided by the sum of true positives and false negatives (asteroid objects marked incorrectly by the system as non-asteroid objects)

$$Recall = \frac{TP}{TP+TN} \quad (4)$$

The F1 score combines the accuracy and recall metrics into one metric. The F1 score has also been developed to perform effectively on data that is unbalanced. If precision and recall values are both high, a model has a high F1 score

$$F1_{score} = 2 * \frac{Precision*Recall}{Precision+Recall} \quad (5)$$

The Receiver Operating Characteristic (ROC) curve is a graphical representation commonly used in binary classification to evaluate the performance of a classifier or a diagnostic test. It illustrates the trade-off between sensitivity (true positive rate) and specificity (true negative rate) across different threshold settings.

IV. RESULTS AND VISUALIZATION

We deployed on the machine leaning methods based on Scikit-learn library, while the deep learning methods based on Tensorflow frameworks. After the preprocessing phase, the datasets will be split into a 80:20 ratio for training and testing, respectively.

We ran it on a computer with 8CPU, RAM 32GB and GPU 8GB on Windows operation. All source code can download at https://github.com/Alvin-Buana/Asteroid-Classification-Benchmark.

*A. The Machine Learning methods*

In the machine learning methods, we evaluate by using F1 score and ROC score. Additionally, we measure the speed of each approach by recording the start and end runtime when applied algorithm to each dataset. The results of F1 and ROC illustrates in Table II.

TABLE II. SCORES ON NEOWS AND ASTEROIDS DATASET IN MACHINE LEARNING METHODS

| Methods | Dataset | | | | | |
| --- | --- | --- | --- | --- | --- | --- |
| | NEOWs | | | Asteroids | | |
| | F1 | ROC | Training time (second) | F1 | ROC | Training time (second) |
| Logistic Regression | 0.94 | 0.93 | 0.03 | 0.98 | 0.98 | 0.01 |
| KNN | 0.90 | 0.90 | 0 | 0.98 | 0.98 | 0 |
| SVC | 0.95 | 0.95 | 6.76 | 0.99 | 0.99 | 45.72 |
| Random Forest | 0.99 | 0.99 | 0.74 | 0.99 | 0.99 | 0.61 |
| CatBoost | 0.99 | 0.99 | 13 | 0.99 | 0.99 | 19.31 |

The Fig.9 and Fig.10 describe the results when we implemented on the NEOWs and the Asteroid datasets, respectively. The both figures show that the Random Forest and Catboost algorithms can get the highest values but the training time of Random Forest algorithm is better than the Catboost method. So that the Random Forest algorithm is the best choice for the asteroid classification.

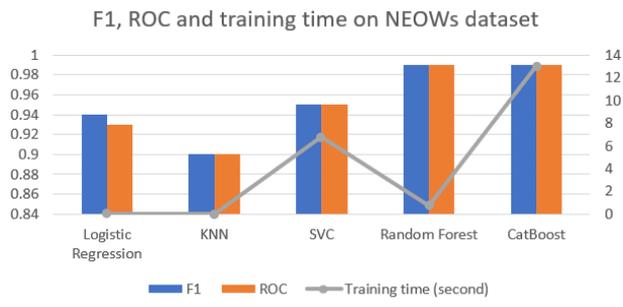

Fig. 9. The F1, ROC and training time on ML methods

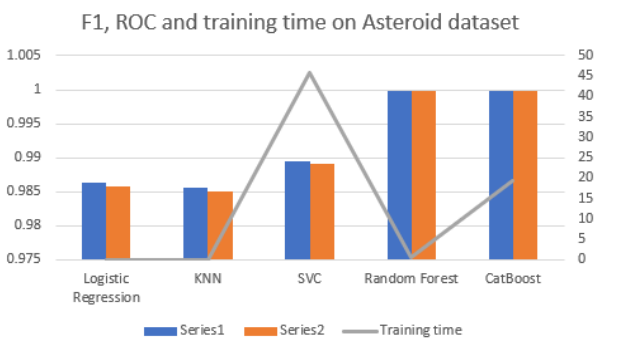

Fig. 10. The F1, ROC and training time on ML methods

## B. The Deep Learning methods

In the Deep Learning methods, we implemented on 5 methods: MPL, DNN, Conv2D, RNN, and LSTM. We also ran it on both datasets: NEOWs and Asteroids.

On the MPL method, we transformed the data and use scikit-learn library to run it with max_iter=300. On the DNN and Conv2D, the dimension of input shape is number of features of dataset. In RNN and LSTM, the data will be modified to the time series data with the number of steps is 1. Similar with Machine Learning methods, in the Deep Learning methods, we use F1 and ROC to evaluate the results of each method, which illustrated in Table III. We ran with 100 epochs for each methodology. The accuracy of each method on NEOWs and Asteroid dataset is presented on the Fig. 11.

TABLE III. SCORES ON NEOWS AND ASTEROIDS DATASET IN DEEP LEARNING METHODS

| Methods | Datasets | | | |
|---|---|---|---|---|
| | NEOWs | | Asteroids | |
| | Training accuracy | Validation accuracy | Training accuracy | Validation accuracy |
| MPL | 0.943 | 0.944 | 0.988 | 0.988 |
| DNN | 0.989 | 0.989 | 0.986 | 0.986 |
| Conv2D | 0.938 | 0.935 | 0.985 | 0.985 |
| RNN | 0.959 | 0.958 | 0.968 | 0.969 |
| LSTM | 0.969 | 0.969 | 0.985 | 0.986 |

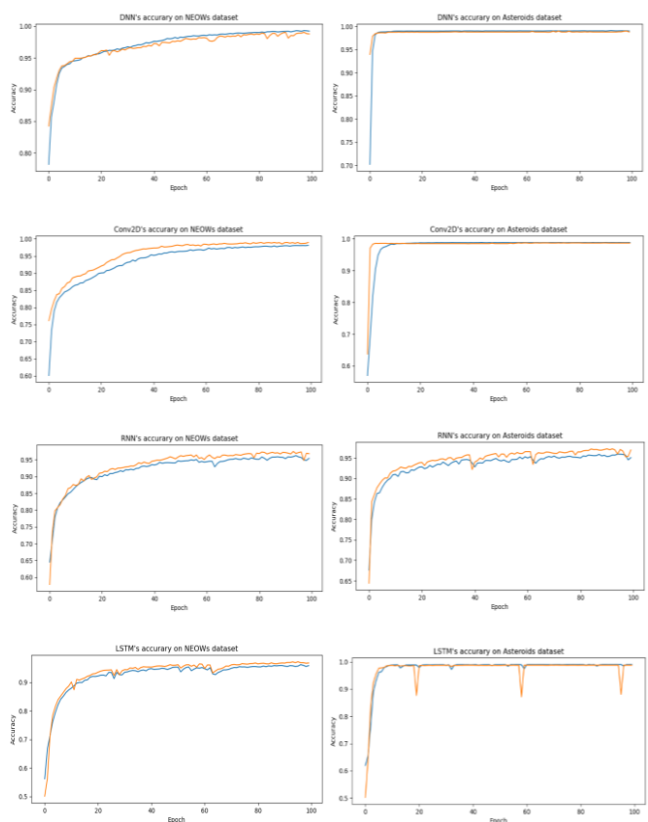

Fig. 11. The accuracy of Deep Learning methods on the datasets

Fig.11 shows that the Deep Learning methods accuracy can increased steadily from 0.5 to almost 1 with 100 epochs, although on RNN and LSTM method it sometimes changes the value downward.

## V. CONCLUSION

We have created a benchmark consist of Machine Learning algorithms and Deep Learning models. In classifying hazardous asteroids, Random Forest performs the best as F1-score and ROC is 0.99. We welcome inputs from the scientific community on how to further improve the performance of these algorithms.

### ACKNOWLEDGMENT

We would like to express our gratitude to Hung-Hsuan Chen for the guidance and learning materials that has been given the Data Science materials and support for our final project.

The Table IV is presented the contributions of the team members.

TABLE IV. CONTRIBUTIONS OF TEAM MEMBER

| Name and student id | Contributions |
|---|---|
| Alvin Buana - 112522608 | Data extraction; data visualization; data preprocessing; model training; writing research methodology report; presentation video. |
| Thai Duy Quy - 111582610 | Model training; result analysis report; data visualization; combining machine learning and deep learning code; proofreading report. |
| Josh Lee- 112522011 | Model training; building presentation; model preliminary explanation; proposal making |
| Rakha Asyrofi- 111582603 | Project planning, assisting proposal |